\documentclass[aps,prl,twocolumn,showpacs,amssymb,amsfonts]{revtex4}
\newcommand{\hil}[1]{\mbox{$\mathcal{#1}$}}
\begin{document}

\title{Generic Incomparability of Infinite-Dimensional Entangled 
States}
\author{Rob Clifton}
\affiliation{Department of Philosophy \\ 1001 Cathedral of
Learning \\ University of
Pittsburgh, Pittsburgh, PA\ 15260  USA \\ E-mail: rclifton@pitt.edu}
\author{Brian Hepburn and Christian W\"uthrich}
\affiliation{Department of History and Philosophy of Science \\ 1017 Cathedral of Learning \\
University of Pittsburgh, Pittsburgh, PA 15260  USA\\
E-mails: brh15@pitt.edu, wuthrich@pitt.edu}
\date{\today}

\begin{abstract} In support of a recent conjecture by Nielsen (1999), we prove that 
the phenomena of `incomparable entanglement'---whereby, neither member of a pair of pure entangled states 
can be transformed into the other via local operations 
and classical communication (LOCC)---is a generic feature 
when the states at issue live in an infinite-dimensional Hilbert 
space.
 \end{abstract} 
\maketitle



\section{Nielsen's Characterization Theorem and Conjecture}   

Let $\hil{H}_{n}$ be a Hilbert space of countable dimension $n\geq 2$.  
  For unit vectors $\psi_{1,2}\in \hil{H}_{n}\otimes\hil{H}_{n}$, 
  i.e., 
  two states of a composite system with two isomorphic subsystems, let 
  $\psi_{1}\prec\psi_{2}$ 
    denote that it is possible to transform 
  $\psi_{1}$ into $\psi_{2}$  with certainty by performing local operations 
  on the subsystems and communicating classically between their 
  locations (LOCC).  (See \cite{nc}, Sec. 12.5.1 for a 
  complete discussion.)  Let $\rho_{\psi_{i}}$ denote the reduced density 
  operator on $\hil{H}_{n}$ determined by the state $\psi_{i}$, and let 
  $\vec{\rho}_{\psi_{i}}=\{\lambda^{(1)}_{i},\ldots,
  \lambda^{(n)}_{i}\}$ denote the vector of  
  $\rho_{\psi_{i}}$'s eigenvalues, i.e., $\psi_{i}$'s squared Schmidt 
  coefficients, arranged in non-increasing order.   Then 
  Nielsen's  \cite{niel} characterization theorem 
  asserts that $\psi_{1}\prec\psi_{2}$ iff $\vec{\rho}_{\psi_{1}}$ is 
  \emph{majorized} by $\vec{\rho}_{\psi_{2}}$, i.e., iff for all $k=1,\ldots,n$, $\sum_{j=1}^{k}\lambda^{(j)}_{1}\leq 
  \sum_{j=1}^{k}\lambda^{(j)}_{2}$.  
  
  One corollary of this 
  elegant little 
  characterization is the following simple result, to be used later on.  The Schmidt number, $\sharp\psi_{i}$, of a state 
  $\psi_{i}$ is defined to be the number of nonzero entries of the 
  vector $\vec{\rho}_{\psi_{i}}$.  Thus, Nielsen's theorem makes it 
  easy to see that a state's Schmidt 
  number cannot be increased under LOCC; for, if 
  $\sharp\psi_{1}<\sharp\psi_{2}$, then the $\sharp\psi_{1}$-th inequality 
  in the majorization condition must necessarily fail due to the 
  normalization of the eigenvalues of a reduced density operator.  
  In particular, then, it follows that a product state, for which the 
  Schmidt number is $1$, cannot be 
  LOCC-transformed into an entangled state---which, of course, we 
  already know must be true, because entanglement between 
  systems cannot be created by local operations on either of them 
  alone. 
    
  Consider, now, the set, $S_{inc}$, of all pairs $(\psi_{1},\psi_{2})$ 
  such that $\psi_{1}\not\prec\psi_{2}$ \emph{and} 
  $\psi_{2}\not\prec\psi_{1}$, where `$inc$' stands for $incomparable$, 
  in Nielsen's \cite{niel} terminology. 
           In the same paper (cf. \cite{niel}, p. 3), Nielsen gave a heuristic argument 
           for the claim  
           that the probability of picking at random two incomparable 
           states  out of the set of all $n\times n$ entangled pure 
           states---according to the natural, rotationally invariant 
           measure---tends to $1$ as 
           $n\rightarrow\infty$.  If true, this conjecture would appear to
           establish that there is a large variety of different 
           non-interconvertible forms 
           of pure state entanglement encountered as the dimension of a system's 
           state space increases without bound.  
           However, it is not 
           obvious how to complete Nielsen's 
           reasoning with a simple but rigorous argument; e.g., 
           \.{Z}yczkowski \& Bengtsson (see \cite{z}, Sec. IIID) have given 
           another 
           argument based upon geometrical considerations, but it too is 
           no more than heuristic.  
           
           So 
           in this note, we shall focus on rigorously
           establishing an elementary but slightly different result that equally well supports 
           the intuition that the complexity of pure state entanglement increases 
           with dimension: namely, when 
           $n=\infty$, the set of pairs in $S_{inc}$ 
           lie 
           open and dense in the Cartesian product of the unit sphere of 
           $\hil{H}_{n}\otimes\hil{H}_{n}$ 
           with itself.  Here, the physically relevant topology is that induced by the 
           standard Hilbert space norm, which, in particular, 
           guarantees that two pairs 
           of pure 
           states will qualify as close only if they (pairwise) dictate 
           \emph{uniformly} close 
           expectation values for all observables.  Due to the fact that the 
           unit sphere of an infinite-dimensional Hilbert space is not even locally compact, there 
           is no sensible Lebesgue-type measure on the set of pairs of 
           unit vectors taken from 
$\hil{H}_{\infty}\otimes\hil{H}_{\infty}$ (cf. \cite{bour}, p. 241).  
           Thus, the statement 
           that $S_{inc}$ is norm open and dense in 
           the infinite-dimensional case is the strongest 
           statement about the 
           genericity of the set of incomparable states that one can 
           possibly hope to make as the complement of $S_{inc}$ is then `nowhere
dense' and so has measure zero.
           
To put it more plainly, genericity, in this context, amounts to
firstly that within any finite region of the 
$\hil{H}_{\infty}\otimes\hil{H}_{\infty}$ 
space of state
pairs there are
uncountably many pairs that are incomparable. Secondly, since the set of 
comparable pairs is closed (as it's the complement of the open set $S_{inc}$)
then there are comparabale states on the boundary of the set for which an 
approximating incomparable state can be found as close as you like. The 
converse cannot be said of any incomparable state. In this sense we claim 
that incomparability is more common than comparability and hence that the 
former is a generic property. 

Moreover, as will be seen, our method of proof actually establishes that 
densely many of these generically incomparable pairs are 
in fact \emph{strongly} incomparable in the sense of Bandyopadhyay 
\emph{et al.} \cite{band}: i.e., they cannot even be converted into one 
another with the help of an entanglement catalyst, or by performing 
collective local operations on multiple copies of the input state.  
Thus our result actually \emph{strengthens} the intuition behind 
Nielsen's conjecture.

\section{Proof of Generic Incomparability for Infinite-dimensional States}
Let us first establish  that $S_{inc}$ is open---or, equivalently, its 
complement $S_{inc}'$ is closed---which happens to be true for \emph{any} 
countable value of $n$.  
           To this end, let us write, when $k$ is finite, `$\psi_{1}\prec_{k}\psi_{2}$' just in case 
           $\sum_{j=1}^{k}(\lambda^{(j)}_{1}-\lambda^{(j)}_{2})\leq 0$; and 
           let us define 
  $S_{\prec_{k}}\equiv\{(\psi_{1},\psi_{2}):
        \psi_{1}\prec_{k}\psi_{2}\}$, with a similar definition 
        for $S_{\succ_{k}}$.   By Nielsen's theorem, 
  $S_{inc}'=(\bigcap_{k=1}^{\infty}S_{\prec_{k}})\cup 
  (\bigcap_{k=1}^{\infty}S_{\succ_{k}})$, and so it suffices 
  for us to show that each 
  $S_{\prec_{k}}$ is closed (the argument for each $S_{\succ_{k}}$ 
  being closed is the same, by 
  symmetry).  
  
  First, note that the mappings $\psi_{i}\mapsto\rho_{\psi_{i}}$ and 
  $\rho_{\psi_{i}}\mapsto\{\lambda^{(1)}_{i},\ldots,
  \lambda^{(k)}_{i}\}$ are 
  both trace-norm continuous (see \cite{clif}, eqns. (1)--(5)), and 
  it is easy to see that 
  the mapping from $\mathbb{R}\mbox{$^{k}$}\times \mathbb{R}\mbox{$^{k}$}$ to $\mathbb{R}$:
  \[ (\{\lambda^{(1)}_{1},\ldots,
  \lambda^{(k)}_{1}\},\ \{\lambda^{(1)}_{2},\ldots,
   \lambda^{(k)}_{2}\})\mapsto\sum_{j=1}^{k}(\lambda^{(j)}_{1}-\lambda^{(j)}_{2})\]
  is jointly continuous. Therefore,  so too is the 
  mapping defined by $\Phi(\psi_{1},\psi_{2})=
  \sum_{j=1}^{k}(\lambda^{(j)}_{1}-\lambda^{(j)}_{2})$.  Now, let 
  $(\psi_{1m},\psi_{2m})\in S_{\prec_{k}} $ be any Cauchy sequence, 
where, by the completeness of Hilbert space, we know there exists a 
limit pair $(\tilde\psi_{1},\tilde\psi_{2})$.  To show that $S_{\prec_{k}}$ is 
closed, we must show that it also contains this limit pair.    But, 
recalling that $\Phi$ is continuous, we know that 
$\{\Phi(\psi_{1m},\psi_{2m})\}=\{\sum_{j=1}^{k}(\lambda^{(j)}_{1m}-\lambda^{(j)}_{2m})\}$ 
must be
 a 
Cauchy sequence too; and, since the nonpositive real numbers are 
closed, it follows that this latter sequence converges to a 
real number 
$\sum_{j=1}^{k}(\tilde\lambda^{(j)}_{1}-\tilde\lambda^{(j)}_{2})\leq 
0$.  Thus $(\tilde\psi_{1},\tilde\psi_{2})\in S_{\prec_{k}}$, as 
required.         
  
    Turning now to the density of $S_{inc}$, what we shall actually 
  establish is that the set of $strongly\ incomparable$ pairs, $S_{st\ 
  inc}$, is dense when $n=\infty$. 
  A pair of entangled states $(\psi_{1},\psi_{2})$ is 
  called \emph{strongly} incomparable just in 
  case $\psi_{1}\not\prec\psi_{2}$ \emph{and} 
  $\psi_{2}\not\prec\psi_{1}$ \emph{and} it is not possible to convert 
  finitely many copies of one of $(\psi_{1},\psi_{2})$ into the other, even with the help of a 
  (finite-dimensional) catalyst.
  To say that $\psi_{1}$ cannot be converted into $\psi_{2}$, even using 
  multiple copies, is to say that for \emph{no} (finite) value of $m$ is it 
  the case that
  \[\underbrace{\psi_{1}\otimes\ldots\otimes\psi_{1}}_{m\ 
  \mbox{times}}\prec \underbrace{\psi_{2}\otimes\ldots\otimes\psi_{2}}_{m\ 
  \mbox{times}}.\] 
   That there are states that cannot be transformed \emph{singly} into each other by 
  LOCC, but \emph{can} be so transformed by local collective operations if multiple 
  copies of the input state are available, was confirmed recently by Bandyopadhyay 
           \emph{et 
           al.} \cite{band}.  To say that a state $\psi_{1}$ cannot be converted 
           into $\psi_{2}$ even with the help of a catalyst is simply to 
           say that there is \emph{no} entangled state $v$ (with finite 
           Schmidt number) such that $\psi_{1}\otimes v\prec\psi_{2}\otimes v$.  
           Again, it was pointed out by Jonathan and Plenio 
           \cite{jp} that there are states that cannot be transformed into each other by 
  LOCC, but \emph{can} be so transformed with the help of a suitable 
  catalyst.
  
  Henceforth, we shall require only one simple \emph{sufficient} condition 
  for a pair of states $(\psi_{1},\psi_{2})$ with finite Schmidt 
  numbers to be strongly incomparable, viz., 
  \begin{eqnarray} \lambda^{(1)}_{1}>\lambda^{(1)}_{2}\ \mbox{AND}\ 
  \sharp\psi_{1}>\sharp\psi_{2}, & \nonumber\\ 
  \mbox{OR}\ \ \ \ \ \ \ \ \ \ \ \ \ \ \ \  & \ \ \ \ \ \ \ \  \mbox{(C)}  \nonumber\\ 
  \lambda^{(1)}_{1}<\lambda^{(1)}_{2}\ \mbox{AND}\ 
  \sharp\psi_{1}<\sharp\psi_{2}.\nonumber\end{eqnarray}
    Let us prove by reductio that this condition, (C), is indeed 
    sufficient for incomparability.  Thus, suppose 
  that, in fact, $(\psi_{1},\psi_{2})$ are \emph{not} strongly 
  incomparable, but that their respective Schmidt values meet 
  condition (C). Then, either for some finite $m$ there is a catalyst $v$ 
  such that $\psi_{1}^{\otimes m}\otimes v\prec \psi_{2}^{\otimes m}\otimes 
  v$ or, similarly, in the reverse direction.   But then, by the first 
  majorization condition in Nielsen's characterization theorem, plus 
  its corollary that a state's Schmidt number cannot be increased under LOCC, it 
  follows that either \[(\lambda^{(1)}_{1})^{m}\lambda^{(1)}_{v}\leq 
  (\lambda^{(1)}_{2})^{m}\lambda^{(1)}_{v}\ \mbox{AND}\ (\sharp\psi_{1})^{m}
  (\sharp v)\geq (\sharp\psi_{2})^{m}
  (\sharp v),\] or that the same two expressions hold with the inequalities 
  reversed.  Thus, upon cancellation, we see that it must be the case that either $\lambda^{(1)}_{1}\leq 
  \lambda^{(1)}_{2}$ and $\sharp\psi_{1}
  \geq \sharp\psi_{2}$, or $\lambda^{(1)}_{1}\geq 
  \lambda^{(1)}_{2}$ and $\sharp\psi_{1}
  \leq \sharp\psi_{2}$---a condition that is easily seen  to be 
 logically inconsistent with (C). 
  
   Turning, finally, to the proof that $S_{st\ inc}$ is dense, first 
   observe
  that the set of all $\psi\in\hil{H}_{\infty}\otimes\hil{H}_{\infty}$ 
  for which the entries of $\vec{\rho}_{\psi}$ are all nonzero---for simplicity, 
  we call these
   \emph{complete} states---is itself 
  dense.   For, if a state $\psi$ merely has a Schmidt decomposition involving 
  finitely many terms, i.e., 
  $\psi=\sum_{j=1}^{p<\infty}\sqrt{\lambda^{(j)}}x^{(j)}\otimes y^{(j)}$, it is 
  approximated arbitrarily closely by the sequence of (normalized) complete states 
  $\psi_{m}=\sum_{j=1}^{\infty}\sqrt{\tilde{\lambda}^{(j)}}\tilde 
  x^{(j)}\otimes \tilde y^{(j)}$ 
  where 
  \[\tilde{\lambda}^{(j)}\equiv\left\{ \begin{array}{ll}
    \lambda^{(j)}/(1+m^{-1})>0 & \mbox{when}\ j\leq p, \\ 
  1/(2^{j-p}(m+1))>0& \mbox{when}\ j>p, \end{array} \right. \]
  and the orthonormal 
  \emph{bases} $\tilde 
  x^{(j)},\tilde 
  y^{(j)}$ respectively extend the orthonormal sets $x^{(j)}, 
  y^{(j)}$ beyond the index value $p$.   Thus, the 
  set consisting of complete pairs of states $(\psi_{1},\psi_{2})$ 
  is a dense set.  Furthermore, it is quite easy to see that any 
  complete pair of states with $\lambda^{(1)}_{1}=
  \lambda^{(1)}_{2}$ can be arbitrarily closely approximated by 
  complete pairs that do \emph{not} satisfy that identity.  
  So, in sum: the set, call it $S_{c_{\not=}}$, of all complete pairs of 
  states, whose first 
  Schmidt coefficients are \emph{un}equal, form a dense set.   We are 
  going to show that every element of $S_{c_{\not=}}$ can itself be 
  approximated arbitrarily closely using members 
  of $S_{st\ inc}$, which therefore must \emph{also} be a dense 
  set.  
  
  So let $(\psi_{1},\psi_{2})\in S_{c_{\not=}}$ be arbitrary.  If $\lambda^{(1)}_{1}>
  \lambda^{(1)}_{2}$, let us choose the sequence of (finite Schmidt 
  number) pairs 
  $(\psi_{1m},\psi_{2m})$ in such a way that
  \[\vec{\rho}_{\psi_{1m}}=
  \left\{\frac{\lambda^{(1)}_{1}}{\sum_{j=1}^{m}\lambda^{(j)}_{1}},\ldots,
    \frac{\lambda^{(m)}_{1}}{\sum_{j=1}^{m}\lambda^{(j)}_{1}},0,0,0,\ldots\right\},\] 
  \[\vec{\rho}_{\psi_{2m}}=
  \left\{\frac{\lambda^{(1)}_{2}}{\sum_{j=1}^{m-1}\lambda^{(j)}_{2}},\ldots,
    \frac{\lambda^{(m-1)}_{2}}{\sum_{j=1}^{m-1}\lambda^{(j)}_{2}},0,0,0,0,\ldots\right\}.\]
  By construction, 
  $\lim_{m\rightarrow\infty}(\psi_{1m},\psi_{2m})
  =(\psi_{1},\psi_{2})$, $\sharp\psi_{1m}>\sharp\psi_{2m}$ for all 
  $m$ (since, by completeness of $(\psi_{1},\psi_{2})$, $\lambda^{(j)}_{1},
  \lambda^{(j)}_{2}\not=0$ for all $j$), 
 and \emph{for all sufficiently large} $m$, $\lambda^{(1)}_{1m}>
  \lambda^{(1)}_{2m}$ (since, $\lambda^{(1)}_{1}>
  \lambda^{(1)}_{2}$).    Thus, the pairs $(\psi_{1m},\psi_{2m})$ 
  approximate $(\psi_{1},\psi_{2})$ and, in virtue of satisfying 
  condition (C), are strongly incomparable for 
  all sufficiently large $m$.  Similarly, if instead $\lambda^{(1)}_{1}<
  \lambda^{(1)}_{2}$ holds for the pair $(\psi_{1},\psi_{2})$, an analogous approximating sequence 
  of (for all sufficiently large $m$, strongly incomparable) pairs  
  $(\psi_{1m},\psi_{2m})$ is obtained simply by interchanging the definitions 
  of $\psi_{1m}$ and $\psi_{2m}$ above.



\begin{thebibliography}{99}

\bibitem{nc} Nielsen, M. A. \& Chuang, I. L. (2000), \emph{Quantum 
Computation and Quantum Information}.  Cambridge: Cambridge 
University Press.

\bibitem{niel} Nielsen, M. A. (1999), `Conditions for a 
class of entanglement transformations', \emph{Physical Review Letters} 
\textbf{83} (2): 436--439.

\bibitem{z} \.{Z}yczkowski, K. \& Bengtsson, I. (2001), `Relativity of 
Pure States Entanglement', {\tt quant-ph/0103027 v2}.

\bibitem{bour} Bourbaki, N. (1984), \emph{Elements of the History of 
Mathematics}.  Berlin: Springer-Verlag.

\bibitem{band} Bandyopadhyay, S., Roychowdhury, V. P., \& Sen, U. (2001). 
`A Classification of Incomparable States',  {\tt quant-ph/0103131}.

\bibitem{clif} Clifton, R. \& Halvorson, H. (1999), `Bipartite Mixed 
States of Infinite-Dimensional Systems are Generically Nonseparable', 
\emph{Physical Review A} \textbf{61}: 012108.

\bibitem{jp} Jonathan, D. \& Plenio, M. B. (1999), 
`Entanglement-Assisted Local Manipulation of Pure Quantum States', 
\emph{Physical Review Letters} \textbf{83}: 3566--3569.






\end{thebibliography}
\end{document}